# Area Versus Speed Trade-off Analysis of a WiMAX Deinterleaver Circuit Design

Omar Rafique

(Dept. of Electronics and Communication, IUST, Awantipora, J & K, India)

*Abstract -* **Trade-off is one of the main design parameters in the field of electronic circuit design. Whereas smaller electronics devices which use less hardware due to techniques like hardware multiplexing or due to smaller devices created due to techniques developed by nanotechnology and MEMS, are more appealing, a trade-off between area, power and speed is inevitable. This paper analyses the trade-off in the design of WiMAX deinterleaver. The main aim is to reduce the hardware utilization in a deinterleaver but speed and power consumption are important parameters which cannot be overlooked.**

## I. INTRODUCTION

WiMAX (Worldwide Interoperability for Microwave Access), defined by IEEE 802.16e standard, was created in 2001 and is capable of delivering up to 70 megabit data rates per second with the added advantage of mobility. The underlying mathematical equations which define the deinterleaver are floor function dependent which are hardware inefficient. Many attempts have been made to make the deinteleaver circuit hardware efficient but the field is still open to research and analysis.

## II. THE WiMAX MODEL

The WiMAX model mainly consists of the following blocks:

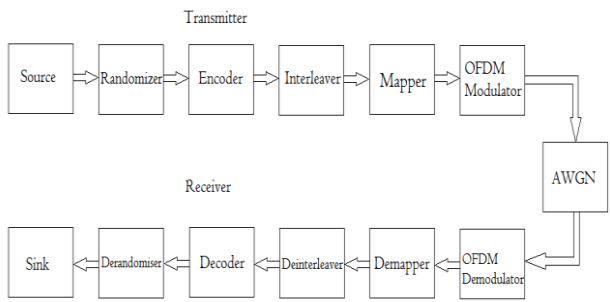

Figure 1 : WiMAX Block Diagram

The randomizer eliminates a long sequence of zeros and ones so that synchronization is not lost. It works on bit by bit basis. Encoding is used for forward error correction where additional redundancy bits are added to the output of the randomizer. Interleaver is used for protection against burst errors which can make a sequence of consecutive bits erroneous, thus making it difficult for the error correcting codes to correct this long sequence of consecutive errors. WiMAX uses Reed Solomon Codes and its error correcting capacity is 8 bits. If there are more than 8 consecutive bits in error than RS code will not be able to correct them. It is the job of the interleaver to break this sequence of consecutive erroneous bits and make it possible to correct errors by RS codes. The mapper maps the incoming bits onto a constellation. The OFDM modulator is an Inverse Fast Fourier Transform device which converts its essentionally digital inputs into analog outputs as the channel is analog in nature and cannot support the transmission of digital bits as such. Therefore, the digital data has to be converted into an analog waveform before being transferred to the receiver through a channel which in most cases is free space. WiMAX uses three modulation techniques in general. They are:

1)     Quadrature Phase Shift Keying(QPSK)

2)     16 – Quadrature Amplitude Modulation (16-QAM).

3)     64 – Quadrature Amplitude Modulation (64 - QAM).

The governing equation for WiMAX interleaver is a two-step permutation and is defined by IEEE 802.16 standard is given by:

$$m_k = \frac{N_{cbps}}{d} \times (k\%d) + \left\lfloor \frac{k}{d} \right\rfloor \qquad (1)$$

$$j_k = s \times \left\lfloor \frac{m_k}{s} \right\rfloor + \left( \left( m_k + N_{cbps} - \left\lfloor d \times \frac{m_k}{N_{cbps}} \right\rfloor \right) \% s \right) \qquad (2)$$

The deinterleaver performs inverse operations and the permutations for it are defined as follows:

$$m_j = s \times \left\lfloor \frac{j}{s} \right\rfloor + \left( j + \left\lfloor \frac{d.j}{N_{cbps}} \right\rfloor \right) \% s \qquad (3)$$

$$k_j = d.m_j - (N_{cbps} - 1) \times \left\lfloor \frac{d.m_j}{N_{cbps}} \right\rfloor \qquad (4)$$

Equations (3) and (4) define a two level permutation and j is the index of the received bit within a block of $N_{cpbs}$ bits. The first equation is for mapping the adjacent codded bits onto non adjacent subcarriers of the OFDM modulation scheme and second equation and the second permutation maps them alternately onto less or more significant bits of the constellation thus avoiding long runs of lowly reliable bits. The letter 'd' represents the number of columns of the block interleaver and may be chosen as 12 or 16. We have preferred d = 16 in this paper as 16 is a power of 2 and division with any number other than the power of 2 is not feasible for FPGA implementation. 'k' varies from 0 to N and $M_k$ is the output of the first equation and $J_k$ is the output of the second equation. The parameter 's' is defined as 1,2 and 3 for QPSK, 16-QAM and 64-QAM respectively, where $N_{cbps}$ stands for numner of codded bits per symbol of the OFDMA modulation scheme. The floor fuction has been represented as $\lfloor \ \rfloor$.

### III. EXISTING METHODS

Two prominent sources are Upadhyaya and Sanyal [1] and Asghar and Liu [2]. [2] has broken down the complex hardware inefficient mathematical equations into simpler equations which do not contain floor functions but still suffers from some mathematical complexities and is not clearly explained whereas [1], though being a simple approach can be further simplified to incorporate random code rates for future developments and at the same time incorporating further hardware efficiency. [3] has tried to overcome the difficulties in [1] and has been able to achieve further hardware efficiency but no trade off analysis has been done in any of the above works.

In this paper our earlier work [3] has been optimized separately with respect to speed and area and it has been shown that for the WiMAX deinterleaver the speed optimized circuit is a better choice than the area optimized circuit due to reasons discussed later on in this paper. Comparison of existing techniques and the speed and area optimized variants of [3] have been done to prove that speed optimized [3] offers the best trade-off between speed and area whereas power consumption is same for both.

### III. TRADE-OFF ANALYSIS:

The area optimization and speed optimization of our earlier work [3] was carried out and the results are in TABLE IV and TABLE V respectively. We used Xilinx ISE Design Suit to develop the project with the following settings.

TABLE I.
PROJECT SETTINGS

| Family | Spartan 3 |
|---|---|
| Device | XC3S400 |
| Package | PQ208 |
| Speed | -5 |

For the area optimized circuit the timing summary and power consumption are as given below:

TABLE II.
AREA OPTIMISED DESIGN SUMMARY FOR TIMING AND POWER

| Timing Summary | Max. Frequency = 107.41 MHz |
|---|---|
| Power Consumption | 56mW |

For the speed optimized circuit the timing summary and power consumption are as given below:

TABLE III.
SPEED OPTIMISED DESIGN SUMMARY FOR TIMING AND POWER

| Timing Summary | Max. Frequency = 130. 2 MHz |
|---|---|
| Power Consumption | 56mW |

It can be deduced from the above tables and figures 2 and 3 that for area optimized design the hardware is decreased by one flip flop and four input LUTs with respect to the speed optimized design, however the frequency is less by an amount of 23 MHz (20 %). This analysis is done at a constant power for both area and speed optimization designs.

TABLE IV:
AREA OPTIMISED DESIGN SUMMARY FOR HARDWARE

| Device Utilization Summary (estimated values) | | | [-] |
|---|---|---|---|
| Logic Utilization | Used | Available | Utilization |
| Number of Slices | 65 | 3584 | 1% |
| Number of Slice Flip Flops | 15 | 7168 | 0% |
| Number of 4 input LUTs | 116 | 7168 | 1% |
| Number of bonded IOBs | 27 | 141 | 19% |
| Number of MULT18X18s | 1 | 16 | 6% |
| Number of GCLKs | 1 | 8 | 12% |

TABLE V:
SPEED OPTIMISED DESIGN SUMMARY FOR HARDWARE

| Device Utilization Summary (estimated values) | | | [-] |
|---|---|---|---|
| Logic Utilization | Used | Available | Utilization |
| Number of Slices | 65 | 3584 | 1% |
| Number of Slice Flip Flops | 16 | 7168 | 0% |
| Number of 4 input LUTs | 120 | 7168 | 1% |
| Number of bonded IOBs | 27 | 141 | 19% |
| Number of MULT18X18s | 1 | 16 | 6% |
| Number of GCLKs | 1 | 8 | 12% |

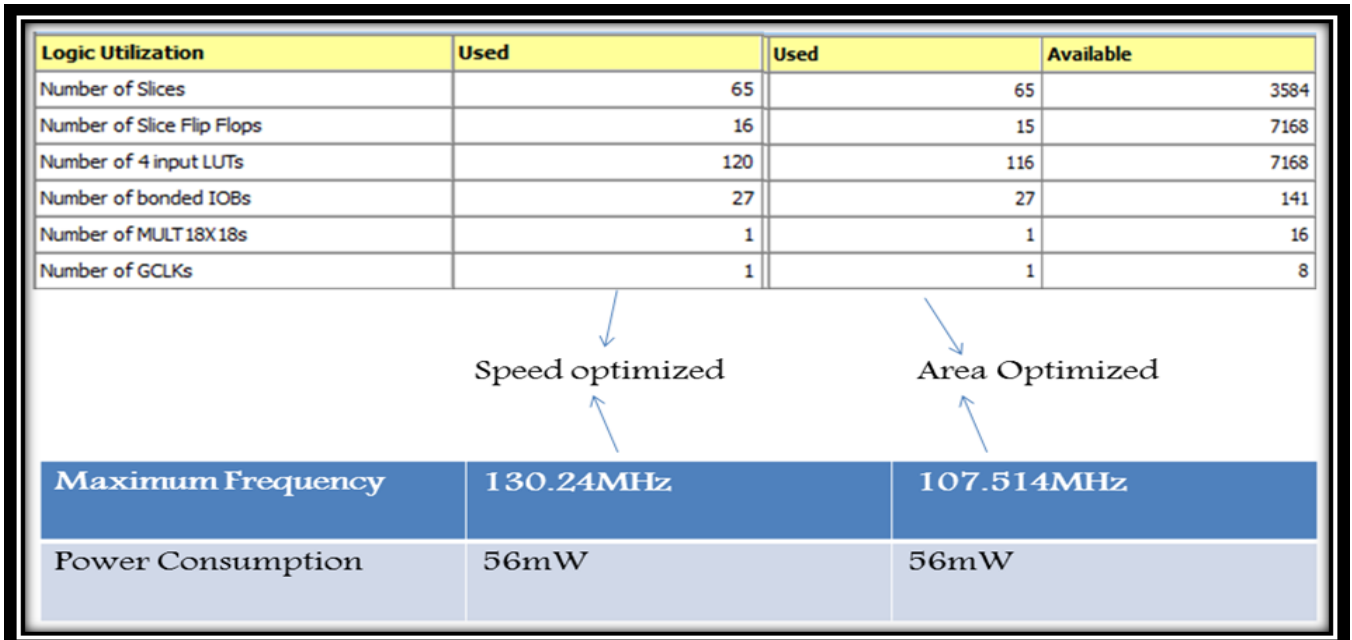

Figure II: Trade Off Analysis

## IV. CONCLUSION

From the analysis in this paper it can be established that even though the hardware consumption of a hardware optimised design for a WiMAX deinterleaver is less, we can observe that it is not a significant improvement when compared to the speed optimised version of the same design. Speed optimised circuit is preferable because it is considerably faster than the area optimised design and when speaking in terms of a trade off we should prefer the speed optimised version. The comparison of the speed optimised version with the existing techniques is given in TABLE VI.

TABLE VI.
COMPARISON BETWEEN SPEED OPTIMISED VERSION OF [3] AND OTHER TECHNIQUES

| Device Specification | Speed optimised version of [3] | Results of Upadhyaya & Sanyal | Performance of LUT based technique | % Reduction w.r.t. Upadhyaya & Sanyal | Remarks |
|---|---|---|---|---|---|
| **Slices** | 1% | 3.49 % | 17.66% | - 71.34 | Decrease |
| **Flip Flops** | 0.153% | 0.50 % | 0.78% | -69.4 | Decrease |
| **4 Input LUTs** | 1% | 3.35 % | 17.75% | -70.14 | Decrease |
| **Operating Frequency** | 130.24MHz | 121.82 MHz | 62.51 MHz | +6.9 | Increase |